\documentclass[a4paper,11pt]{article}
\usepackage{pos}

\title{Is PKS\,0625$-$354 another variable TeV active galactic nucleus? }
 \ShortTitle{PKS\,0625$-$354 }

\author*[a]{Dorit Glawion}
\author[b,c]{Alicja Wierzcholska}

\affiliation[a]{Friedrich-Alexander-Universit\"at  Erlangen-N\"urnberg,  Erlangen  Centre  for  Astroparticle  Physics, Erwin-Rommel-Str. 1,  D-91058,  Erlangen,  Germany}

\affiliation[b]{Instytut  Fizyki  Jadrowej  PAN,  ul.   Radzikowskiego  152,  31-342  Krakow,  Poland}

\affiliation[c]{Landessternwarte,  Universit\"at  Heidelberg,  K\"onigstuhl,  D-69117  Heidelberg,  Germany}

\forColl{H.E.S.S.} 

\emailAdd{dorit.glawion@fau.de}
\emailAdd{alicja.wierzcholska@ifj.edu.pl}

\abstract{

The majority of the active galactic nuclei (AGN) detected at very-high-energies above 100\,GeV belong to the class of blazars with a small angle between the jet-axis and the line-of-sight. Only about 10 percent of the gamma-ray AGN are objects with a larger viewing angle resulting in a smaller Doppler boosting of the emission. Originally, it was believed that gamma-ray emission can only be observed from blazars and those are variable in its brightness. Instead, the last years have shown that non-blazar active galaxies also show a fascinating variability behaviour which provide important new insights into the physical processes responsible for the gamma-ray production and especially for flaring events.

Here, we report on the observation of gamma-ray variability of the active galaxy PKS\,0625$-$354 detected with the H.E.S.S. telescopes in November 2018. The classification of PKS\,0625$-$354 is a still matter of debate. The H.E.S.S. measurements were performed as part of a flux observing program and showed in the first night of the observation a detection of the object with $>5\sigma$. A denser observation campaign followed for the next nine nights resulting in a decrease of the gamma-ray flux. Those observations were accompanied with \textit{Swift} in the X-ray and UV/optical band allowing for the reconstruction of a multi-band broad-band spectral energy distribution. We will discuss the implications of the gamma-ray variability of the object.

}

\FullConference{37$^{\rm{th}}$ International Cosmic Ray Conference (ICRC 2021)\\
		July 12th -- 23rd, 2021\\
		Online -- Berlin, Germany}

\begin{document}

\def\pks0625{PKS\,0625$-$354}
\def\apj{ApJ}
\def\apjs{ApJS}
\def\aap{A\&A}
\def\mnras{MNRAS}

\maketitle

\section{Introduction}

\begin{figure}[h!]
   \centering
   \includegraphics[width=7.cm]{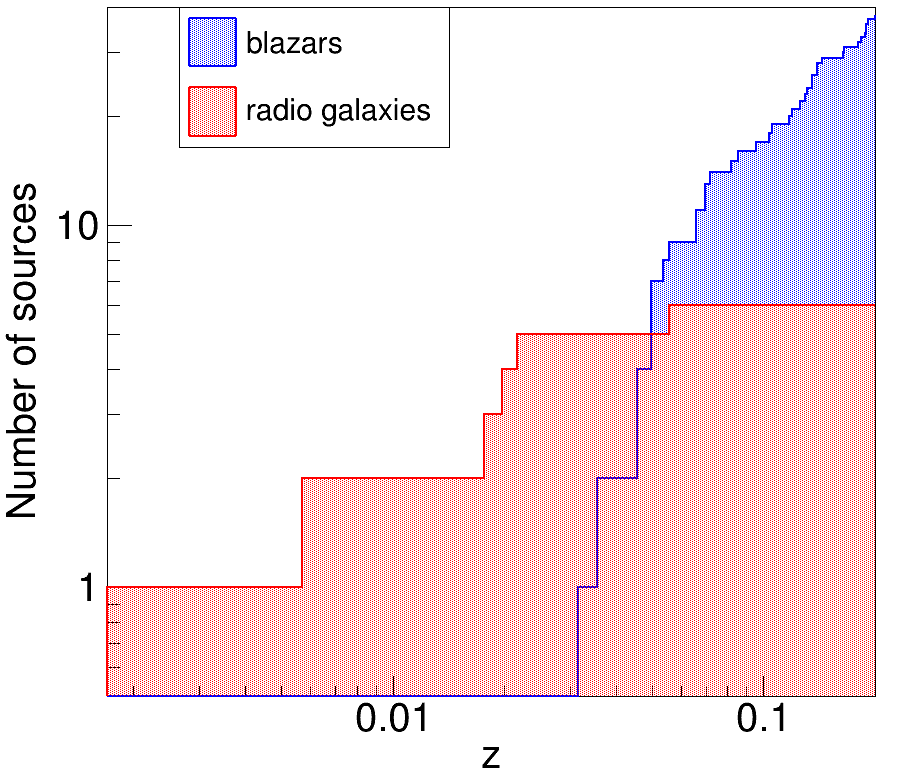}
     \caption{Cumulative number densities of active galaxies emitting TeV radiation within one gigaparsec distance to the Sun. While blazars (blue histogram) are more numerous, radio galaxies (red histogram) are the dominant population in the local universe.
   }
              \label{Fig:Histo}%
    \end{figure}

Out of over 80 AGN detected at TeV energies so far\footnote{\url{http://tevcat.uchicago.edu/}}, only a few are classified as radio galaxies or sometimes called misaligned blazars: 
M87 ($z=0.0044$), Cen\,A ($z=0.002$), NGC\,1275 ($z=0.018$), and very recently also 3C\,264 ($z=0.022$). Furthermore, IC\,310 ($z=0.019$) and PKS\,0625$-354$ ($z=0.055$), classified as radio galaxies in the literature, appear as AGN of unknown type.
Contrary to blazars, radio galaxies are observed at large angle between the line of sight and the jet axis.
On one hand, this results in very weak relativistic beaming making it more difficult to detect distant objects (see Fig.~\ref{Fig:Histo}). 

However, the lack of strong beaming simplifies the interpretation and modeling of the emission of the object, as the spectral energy distribution (SED) of the source is a strong function of the often-unknown Doppler factor. 
Moreover, strong beaming of the jet emission in blazars might hide other emission sources, not connected with the jet, such as the magnetosphere emission or emission from the radio lobes. 
Therefore, such more exotic emission mechanisms are much easier to study in radio galaxies. 

\pks0625 is a Fanaroff-Riley (FR) I radio galaxy \cite{oj10}, also classified as a LINER \cite{le03}. While the optical spectrum ([OIII] line luminosity) resembling that of a BL\,Lac object \cite{wi04}, the kpc radio morphology clearly shows two extended lobe structures typical for a FR~I radio galaxy.
Radio observations revealed superluminal motion with $v_{\mathrm{app}}\propto3.0\pm0.5\,c$ \cite{mu13}.
This corresponds to the limit on the observation angle of the jet $\theta\lesssim37^\circ$.
The jet-to-counterjet ratio of the pc-scale radio jet (one-sided) limits the viewing angle to $\theta\lesssim57^\circ$.
The modeling of the non-simultaneous SED with a synchrotron-self Compton model is consistent in the X-ray to gamma-ray range with the values of the observations angle of $10-19^\circ$ \cite{fu15}.
H.E.S.S. reported on the detection of \pks0625\ in \cite{hess18}.
The observed TeV gamma-ray spectral index of the object is consistent with a simple power-law of the form $\mathrm{d}N/\mathrm{d}E\propto E^{-\Gamma}$ with an index of $\Gamma=2.8\pm0.5$.   
We note that the source is affected by Extragalactic Background Light (EBL) absorption, so the intrinsic spectral index is $\sim 2.5$, similar to that seen in M87 and Cen\,A.
The gamma-ray emission of the object is rather faint. The flux measured by H.E.S.S. above 580\,GeV is $\sim 4\%$ of Crab.
However, taking into account the relatively high redshift of this radio galaxy, it is among the brightest absolute luminosities from the VHE-detected radio galaxies. 
No variability of the very-high-energy emission was detected by H.E.S.S., but observation lasted just 5.5 hours \cite{hess18}.
The average spectral index of \pks0625 in 3FGL catalog is $1.89$. However, a curvature is present \cite{ac15}.
The X-ray emission of \pks0625 is dominated by the jet \cite{fu15}.
Observations with Suzaku show weak (10\% amplitude) X-ray variability on time scales of 1-2 days, while the Fermi-LAT observations revealed a flare in gamma-rays with a rather weak amplitude change by a factor of three \cite{fu15}. The multi-wavelength light curve shows variability in the X-ray band, but no variable emission in the Optical-UV regime \cite{hess18}.

The search for fast variability from misaligned blazars is of particular importance because it can hardly be explained by the standard shock-in-jet model. In small emission region (as inferred from the variability time scale) $\gamma-\gamma$ pair production would lead to the absorption of the TeV gamma-rays making the emission
impossible to be observed. This indicates highly anisotropic radiation processes (to avoid pair absorption). Possible explanations are, e.g., mini-jet structures within the jets \cite{giannios}, jet-cloud/star interactions where the clouds may originate from stellar winds \cite{barkov}, and magnetospheric models \cite{levinson}, similar to those known from pulsar theory.

In order to investigate the variability behaviour of PKS\,0625$-354$ at very-high-energies, H.E.S.S. monitored the emission in November 2018. A significant signal was found in the first night of the observation with the Real-Time-Analysis of H.E.S.S. Therefore, a denser monitoring was organized together with \textit{Swift} observations.

\section{H.E.S.S. observations and analysis results}

In 2018 November from MJD\,58423.99 to MJD\,58432.97 \pks0625 was observed with H.E.S.S. in a hybrid mode with five telescopes in stereoscopic mode. The so-called "wobble mode" was used for the observation where the source position is offset by 0.7$^\circ$ from the camera center. This allows a measurement of the background simultaneously \cite{berge2007}. In the first night of the observation the real-time-analysis running in parallel to the observation indicated a high significance ($\sim 5\sigma$) \cite{balzer}. 
Therefore, further observations were scheduled resulting in a total of $\sim$17.5\,h of live time of good quality data.  

The data were analyzed using the ImPACT maximum likelihood-based fitting technique \cite{ImPACT}. The background produced by cosmic-rays is rejected using a neural network based scheme. The residual background contamination level of the source region (ON and OFF) is estimated with the reflected-background method for the reconstruction of the spectrum and light curve \cite{Crab2006}.  An independent analysis chain was used for the calibration and reconstruction as a crosscheck which yield compatible result. 

The analysis of $\sim$17.5\,h of data using the ring-background method \cite{Crab2006} results in a significance of $8.7\sigma$. The reflected-background method resulted in $8.6\sigma$. 

\begin{figure}
   \centering
   \includegraphics[width=7.cm]{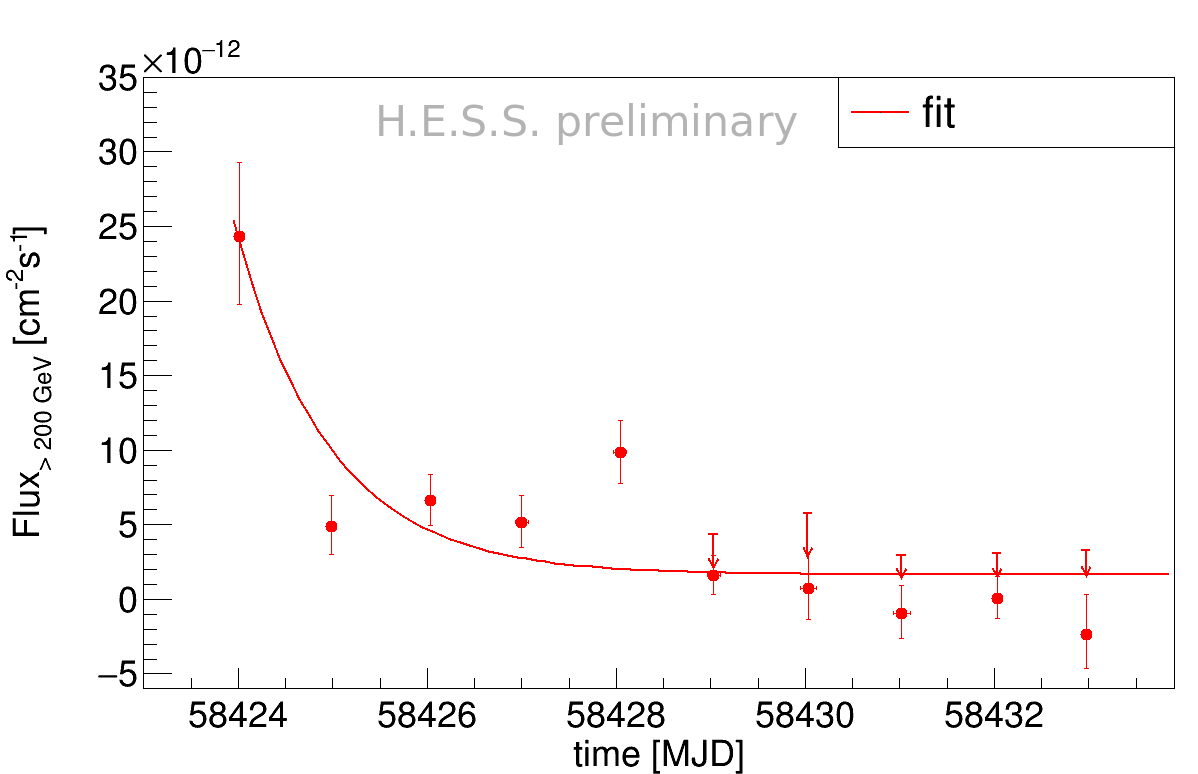}
     \caption{Preliminary nightly binned light curve above 200\,GeV measured with the H.E.S.S. telescopes in November 2018.   }
              \label{Fig:LC}%
    \end{figure}

The nightly-binned light curve above an energy threshold of 200\,GeV is shown in Fig.~\ref{Fig:LC}. Integral upper limits are given for flux data points with less than $2\,\sigma$ and where calculated following \cite{rolke} at the 95\% confidence level. 
Fitting the light curve with a constant fit function yields a $\chi^{2}$ of 55 and 9 degree of freedom (d.o.f.) corresponding to a probability of $1\times10^{-8}$. Furthermore, we fit the light curve with the following function:
\begin{equation}
F=F_0+F_1\cdot2^{-|t-t_1|/\tau},   
\end{equation}
where $\tau$ is the flux doubling time scale. The fit yields a $\tau=(17\pm1)$\,h with a probability of 0.02\%.

\begin{figure}
   \centering
   \includegraphics[width=7.cm]{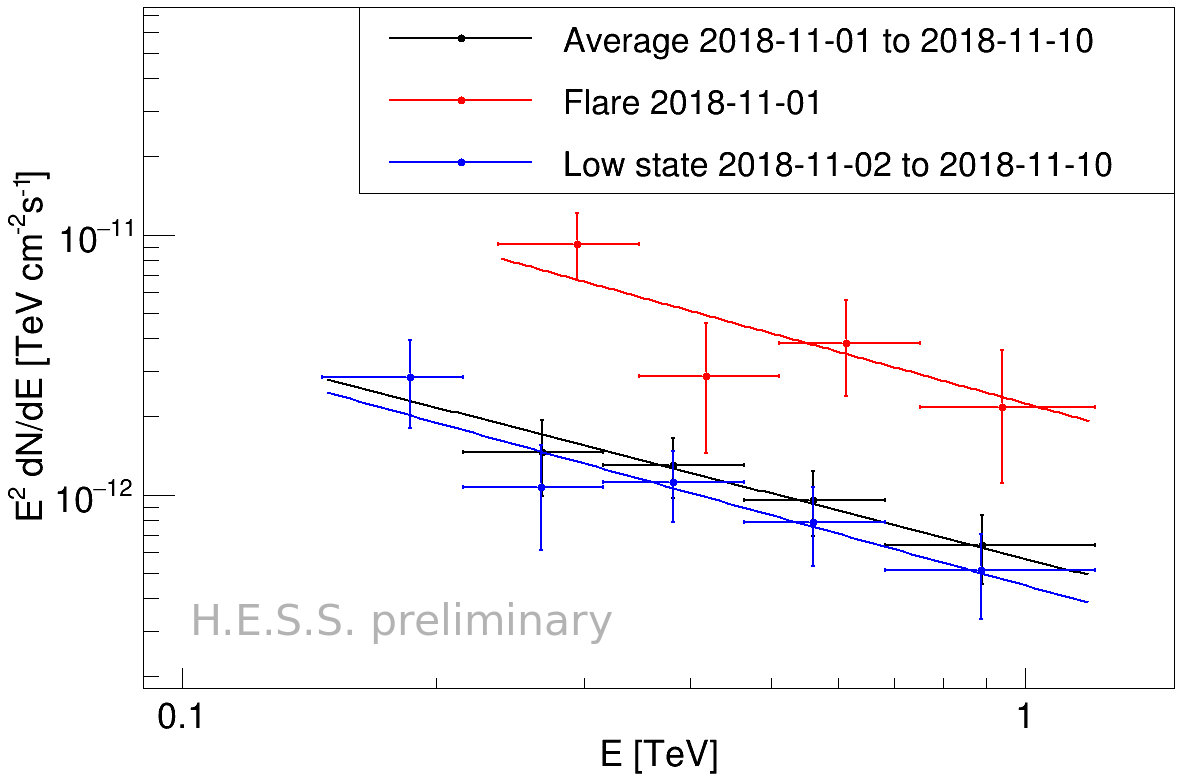}
     \caption{Preliminary spectral energy distributions of different emission states.   }
              \label{Fig:SEDs}%
    \end{figure}

We performed a spectral analysis with a simple power law fit $\mathrm{d}N/\mathrm{d}E=f_0\cdot E^{-\Gamma}$ of all the data (average state), of only the first night (flaring state) and the remaining data (low state). The results are shown in Fig.~\ref{Fig:SEDs} and Table~1.

\begin{table}[h!]
\small
\label{table:3}     
\begin{center}
\begin{tabular}{l l c c c c l}     
\hline
Emission  & dates   & $f_0\times10^{-12}$  & $E_0$  & $\Gamma$  & Energy range &  Notes  \\ 
state	& [h]  & $[\mathrm{TeV}^{-1}\mathrm{s}^{-1}\mathrm{cm}^{-2}]$ & [TeV] &  &  [TeV] &      \\
\hline
Average	& 2018-01-01 - 2018-01-10   & $7.51\pm1.02_\mathrm{stat}$   &  0.40 &  $2.83\pm0.26_\mathrm{stat}$ & 0.15--1.20 &       \\
Flare	& 2018-01-01	&$23.8\pm4.9_\mathrm{stat}$	& 0.44	&$2.90\pm0.49_\mathrm{stat}$	& 0.24--1.20	&	\\  
Low state	&2018-01-02 - 2018-01-10	&$6.27\pm1.02_\mathrm{stat}$ & 0.40	&$2.90\pm0.31_\mathrm{stat}$	& 0.15--1.20 &		\\
2012 	&2012-11 -2012-12 	&$0.58\pm0.22_\mathrm{stat}$	& 1.0	&$2.84\pm0.50_\mathrm{stat}$	& 0.2--10.0 & \cite{hess18}		\\
\hline
\end{tabular}
\end{center}
\caption{Preliminary results of the spectral analysis  of the H.E.S.S. data. Dates are given for the night before sunset. }
\end{table}

 \section{Multi-wavelength spectral energy distribution}
 
\begin{figure}
   \centering
   \includegraphics[width=7.cm]{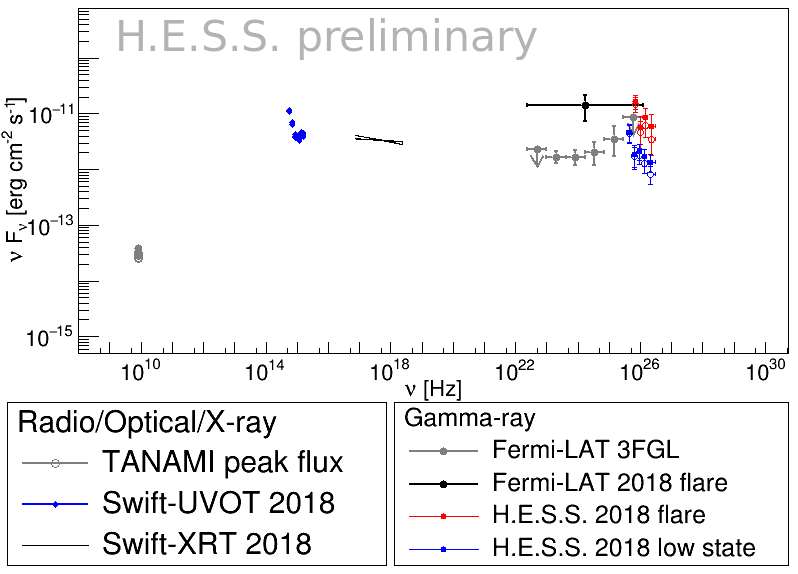}
   \caption{Preliminary multi-wavelength spectral energy distribution.}
              \label{Fig:MWLSEDs}%
    \end{figure}

The H.E.S.S. spectral energy distributions corrected for EBL absorption using the model from \cite{dom} are shown in the multi-wavelength SED in Fig.~\ref{Fig:MWLSEDs}. 

The SED as includes contemporaneous optical/UV and X-ray data obtained with the \textit{Swift}-UVOT and XRT. The X-ray spectrum shown in the figure is the average spectrum obtained from three individual pointings on 2018-11-03, 2018-11-04, and 2018-11-05 showing a simple absorbed power-law with an energy flux density of $F_{2-10\,\mathrm{keV}}=(5.03\pm0.20)\times10^{-12}$\,erg\,cm$^{-2}$\,s$^{-1}$  and a photon spectral index of $2.20\pm0.06$.  In the MeV-GeV band we show the 3FGL catalog spectrum as well as a differential flux point between 100\,MeV--500\,GeV measured with \textit{Fermi}-LAT around the flaring episode from 2018-10-30 to 2018-11-03 with a test statistics of $TS=11$. Note that previously, only a non-simultaneous SED was investigated in \cite{hess18}. Furthermore, we included radio very-long-baseline-interferometry measurements obtained with the TANAMI array \cite{angioni19}.

\section{Implications on viewing angle}

Taking into account the flux-doubling time scale obtained from the TeV light curve we can estimate the size of the emission region.

 \begin{equation}
  R<\delta\cdot c\cdot\tau_\mathrm{var}\cdot\frac{1}{(1+z)},
 \end{equation}
 with $z=0.055$ and $\tau_\mathrm{var}=17$\,h, we obtain
 \begin{equation}
  R<\delta\cdot1.7\times10^{15}\,\mathrm{cm}.
 \end{equation}

We further compare the variability time scale with the light crossing time of the black hole assuming the mass of \pks0625 $M_\mathrm{BH}=(1.55\pm0.66)\times10^{9}M_\odot$ as found by \cite{mass}
\begin{equation}
 t_\mathrm{G}=\frac{GM_\mathrm{BH}}{c^3}=82\,\mathrm{min}.
\end{equation}
Thus, the size of the TeV emission region obtained with the variability time scale is much larger than the radius of black hole assuming that it is maximally rotating. 

Taking into account the small size of the TeV emission region we further calculate the optical depth for  gamma-gamma pair production for 1\,TeV photons using Eq.\,9 in \cite{equation}:
\begin{equation}
 \tau_{\gamma\gamma}\sim\frac{\sigma_\mathrm{T}D_\mathrm{L}^2F_0\epsilon_\gamma(1+z)}{10Rm_e^2c^5\delta^5}<1
\end{equation}
where $\sigma_\mathrm{T}$ is the Thomson cross section, $D_\mathrm{L}$ is the luminosity distance, $F_0$ the peak of the target spectral energy distribution, $\epsilon_\gamma$ the energy of the highest gamma-ray photon (here 1\,TeV), $R$ is the size of the emitting region, $m_e$ the mass ofthe electron and $c$ the speed of light. Assuming $F_\mathrm{\mathrm{K,H bands}}= 1.75\times10^{-11}$\,erg\,cm$^{-2}$\,s$^{-1}$ for $F_0$ where the absorption with the TeV photons is supposed to happen, this yields a constraint for the Doppler factor of $\delta>2.4$.

\begin{figure}
   \centering
   \includegraphics[width=7.cm]{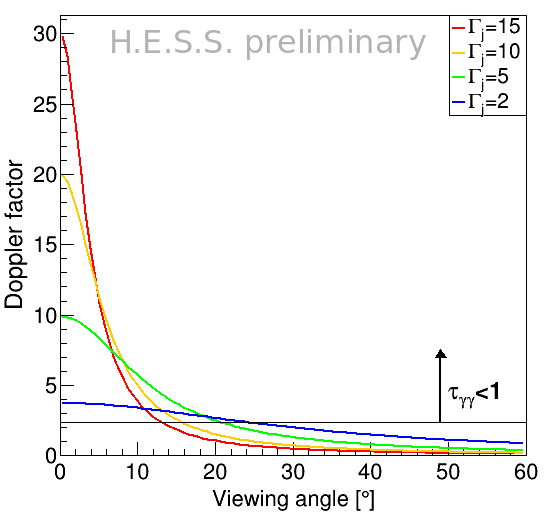}
     \caption{Doppler factor versus viewing angle as a function of the Lorentz factor. The solid black line indicates the doppler factor for which the optical depth for $\gamma\gamma$-pair-production becomes 1.
   }
              \label{Fig:Angle}%
    \end{figure}

In Fig\,~\ref{Fig:Angle} we plot the Doppler factor $\delta=(\Gamma_b(1-\beta\cos\theta))^{-1}$ as a function of the viewing angle $\theta$ assuming different bulk Lorentz factors. In addition we plot the constraint for the Doppler factor of $\delta>2.4$.

Thus, the Doppler factor constraint from the TeV variability provides an upper limit on the viewing angle of  \pks0625 of $\theta<24^\circ$ in comparison to the upper limit of $\theta<53^\circ$, reported previously based on the flux of the jet and the counterjet at parsec-scale radio maps \cite{angioni19}. 
 
\section{Summary and Conclusion}

For the first time we have observed variability in the very-high-energy band from the AGN \pks0625. Furthermore, we presented a contemporaneous measured multi-wavelength SED of the AGN \pks0625. The variability provides a possibility to get an estimate of the upper limit on the viewing angle of the object of $\theta<24^\circ$ which is a tighter constraint than limits from radio measurements. In the future, we will work further on the implications of the viewing angle on the multi-wavelength SED.

\section{ACKNOWLEDGEMENTS}

H.E.S.S. gratefully acknowledges financial support from the agencies and organizations listed at \url{https://www.mpi-hd.mpg.de/hfm/HESS/pages/publications/auxiliary/HESS-Acknowledgements-2019.html}. A.W. is supported by the National Science Center through the research project  UMO-2016/22/M/ST9/00583.

\clearpage
\section*{Full Authors List: \Coll\ Collaboration}

\scriptsize
\noindent
H.~Abdalla$^{1}$, 
F.~Aharonian$^{2,3,4}$, 
F.~Ait~Benkhali$^{3}$, 
E.O.~Ang\"uner$^{5}$, 
C.~Arcaro$^{6}$, 
C.~Armand$^{7}$, 
T.~Armstrong$^{8}$, 
H.~Ashkar$^{9}$, 
M.~Backes$^{1,6}$, 
V.~Baghmanyan$^{10}$
% ...

\medskip

\noindent
$^{1}$University of Namibia, Department of Physics, Private Bag 13301, Windhoek 10005, Namibia\\
$^{2}$Dublin Institute for Advanced Studies, 31 Fitzwilliam Place, Dublin 2, Ireland\\
$^{3}$Max-Planck-Institut f\"ur Kernphysik, P.O. Box 103980, D 69029 Heidelberg, Germany\\
$^{4}$High Energy Astrophysics Laboratory, RAU,  123 Hovsep Emin St  Yerevan 0051, Armenia\\
$^{5}$Aix Marseille Universit\'e, CNRS/IN2P3, CPPM, Marseille, France\\
$^{6}$Centre for Space Research, North-West University, Potchefstroom 2520, South Africa\\
$^{7}$Laboratoire d'Annecy de Physique des Particules, Univ. Grenoble Alpes, Univ. Savoie Mont Blanc, CNRS, LAPP, 74000 Annecy, France\\
$^{8}$University of Oxford, Department of Physics, Denys Wilkinson Building, Keble Road, Oxford OX1 3RH, UK\\
$^{9}$IRFU, CEA, Universit\'e Paris-Saclay, F-91191 Gif-sur-Yvette, France\\
$^{10}$Instytut Fizyki J\c{a}drowej PAN, ul. Radzikowskiego 152, 31-342 Krak{\'o}w, Poland\\
% ...

\end{document}